%% file: coelho-august2013.tex
\documentclass[useAMS,usenatbib]{mn2e}
\usepackage{graphicx}
\usepackage{times}
\title[Red bulgeless galaxies in SDSS DR7. Are there any AGN hosts?]{Red bulgeless galaxies in SDSS DR7. Are there any AGN hosts?}

\author[Coelho et al.]{B. Coelho$^{1,2}$\thanks{email} and S. Ant\'on$^{3,4}$ 
and C. Lobo$^{1,2}$ and B. Ribeiro$^{1,2}$\\
$^{1}$ Centro de Astrof\'isica, Universidade do Porto,
Rua das Estrelas, 4150-762 Porto, Portugal\\ 
$^{2}$ Departamento de F\'isica e Astronomia, Faculdade de Ci\^encias,
Universidade do Porto, Rua do Campo Alegre, 4169-007 Porto, Portugal\\
$^{3}$ CICGE, Faculdade de Ci\^encias da Universidade do Porto, Campo Alegre, Porto, Portugal\\
$^{4}$ SIM, Faculdade de Ci\^encias da Universidade de Lisboa, Campo Grande, Lisboa, Portugal} 

\begin{document}

\date{Received ...; accepted ... ..., ...}

\pagerange{\pageref{firstpage}--\pageref{lastpage}} \pubyear{2013}

\maketitle

\label{firstpage}

\begin{abstract}
With the main goal of finding bulgeless galaxies harbouring
super massive black holes and showing, at most, just residual star formation activity, we
have selected a sample of massive bulgeless red sequence galaxies
from the SDSS-DR7, based on the NYU-VAGC catalogue. Multivavelength
data were retrieved using EURO-VO tools, and the objects are characterised
in terms of degree of star formation and the presence of an AGN.
 We have
found seven objects that are quenched massive galaxies, that have no
prominent bulge and that show signs of extra activity in their 
nuclei, five of them being central in their halo. These objects are rather robust candidates for rare systems that, though devoid of a significant bulge, harbor a supermassive black hole with an activity level likely capable of having halted the star formation through feedback.
\end{abstract}


\section{Introduction}
In the currently most popular paradigm of galaxy formation and evolution, 
massive galaxies develop a significant bulge that harbours a supermassive 
black hole \citep[SMBH; eg.][]{Cattaneo09}.  This central engine may
 exert a feedback - either by 
driving gas out of the galaxy \citep{KauffHaeh00,Hopkins06,Tremonti07} or 
by preventing halo gas from cooling and feeding the disk 
\citep[eg.][]{Croton06,Bower2006} - that quenches star formation and starts moving 
the galaxy towards the present-day well-defined red sequence. Semi-analytic 
models now include this effect \citep[eg.][and references therein]{Croton06,Bower2006}
 to successfully minimise the excess of star-forming galaxies populating 
the bright end of the luminosity function produced otherwise, and to 
reproduce the observed correlation between the masses of the bulges of 
AGN host galaxies and the SMBH they harbour \citep{Gebhardt00,FerraMerr00}.
One of the consequences of the above scenario is that bulgeless galaxies 
quenched via this mechanism are expected to be statistically rare objects and this was
recently confirmed by \citet{Bell08} based on SDSS DR2 derived data.\\
Interestingly enough, recent studies have been
finding an increasing number of bulgeless galaxies harbouring  
 AGNs, with black hole mass estimates ranging 
between  M$_{\mbox{\tiny {BH}}}$ $\sim 10^3$ and $10^7$ M$_\odot$  \citep[][to cite a few]{Satyapal07,Satyapal08,Satyapal09,DesrHo09,Masters2010,McAlpine11,Secrest12,Simmons2013}, showing that a massive bulge does not seem to be a requirement for the presence of a SMBH. Is the latter capable of disrupting star 
formation among bulgeless host galaxies?  Here we try to contribute for 
that discussion through
a multi-wavelength study of a sample of red and bulgeless massive galaxies 
drawn from the SDSS dataset DR7\footnote{http://www.sdss.org/DR7/}, 
compiled in a high quality catalogue: the New-York University Value 
Added Galaxy Catalogue \citep[NYU-VAGC][]{Blanton05}\footnote{http://sdss.physics.nyu.edu/vagc/}. We analyse the properties of the galaxies in our sample in terms of structure, star formation rate, AGN content and central versus satellite status, and discuss our findings in 
the context of AGN-quenched galaxy scenarios.
We adopt H$_0$ = 70 km s$^{-1}$ Mpc$^{-1}$, $\Omega_m$ = 0.3 and 
$\Omega_\Lambda$ = 0.7.

\section{Sample Selection}\label{sec_sample}
 Our investigation is based on the SDSS DR7 dataset, using the NYU-VAGC 
catalogue from  Blanton et al. (2005) that gathers  estimates for structural 
and  photometric parameters for all galaxies of the SDSS having spectroscopic 
data.  The objects were selected according to the following criteria:

\begin{itemize}
\item  Redshift in the range $0.02 < z < 0.06$;
\item  Galaxy stellar mass $M_* > 10^{10} M_\odot$,  as 
 this is the mass range where AGN feedback is expected to be the dominant 
 mechanism affecting star formation \citep{Bell08}; 
\item S\'ersic index $n < 1.5$ in the r-band, to select galaxies without a relevant bulge component;
\item Colour index $(g-r)$ typical of red galaxies:
 $(g-r) > 0.57 + 0.0575 log_{10}(M_*/10^8M\odot)$  \citep{Bell08};
\item Inclination cut equivalent to $i \la 60^o$  to avoid 
edge-on systems, prone to being reddened by dust extinction.
\end{itemize}

\noindent All magnitudes were k-corrected and de-reddened from Galactic foreground extinction 
(with the \citealt{Schlegel98} galactic maps), and were directly available at the NYU-VAGC. Stellar masses assume an initial mass function of Chabrier (2003) and are computed at the expense of a model mag (g-r) colour index and an r-band luminosity obtained from the respective absolute Petrosian magnitude as in Bell (2008). 
The inclination was calculated using the adaptive second order 
moments determined for the  galaxy's image light profile \citep[see eg.][]{Vincent05}: $q_{am} = [(1 - E)/(1 + E)]^{1/2}$, 
where $E = (me1^2 + me2^2)^{1/2}$  with $me1$ and $me2$ being the adaptive 
moments extracted from the r-band light profile 
(directly available in the NYU-VAGC), and setting $q_{am}$ to values $>0.5$ \citep[again, following][]{Vincent05}.\\
\noindent All the objects were thoroughly checked visually and we 
conservatively eliminated all entries that presented {\em any} of the following characteristics: (i) resulted from obvious errors in the SDSS automatic detection procedure or obvious (automatic) miscalculated photometric or structural parameters; (ii) showed any hint of structure identifiable with either star-forming regions or dust lanes; (iii) belonged to interacting systems.
\noindent We stress that this strict selection was adopted so as to eliminate {\em any “contaminant”} from the final sample; the goal is to find robust candidates of massive galaxies with no significant star formation that, though devoid of a significant bulge, may harbour an AGN that may be the responsible for the 
quenching of the star forming activity. At this stage, our sample comprises 43 objects, and is presented in Table \ref{table_sample}.

\section{Data mining}
\label{spectradata}

Multiwavelength data was retrieved, using EURO-VO tools, from several available catalogues - see Table \ref{table_radius}. We adopted a proximity criterion to identify our objects in the various catalogues, where we assumed the NYU-VAGC coordinates and then 
applied a search radius to cross-match with other catalogues.

\begin{table}
\caption[]{\label{table_radius}List of accessed catalogues, bands, search radii and respective references, and  percentage of objects identified.}
\begin{tabular}{llll}
 \hline \hline
 Catalogue & band & search radius & \% obj \\
 \hline
{\scriptsize NVSS}    & 1.4 GHz & 45\arcsec $ $  $^{(a)}$ & 23.3\% \\
{\scriptsize FIRST}   & 1.4 GHz & 5.4\arcsec  $ $  $^{(b)}$  & 18.6\% \\
{\scriptsize IRASfsc} & 12, 25, 60, 100 $\mu$m & 40\arcsec $  ^{(c)}$  &30.2\% \\
{\scriptsize WISE}    & 3.4, 4.6 $\mu$m & 6.1\arcsec, 6.4\arcsec$ ^{(d)}$ & 100\% \\
                      & 12, 22 $\mu$m & 6.5\arcsec, 12.0\arcsec $ ^{(d)}$ & 100\% \\
{\scriptsize 2MASSpsc} & J, H, K & 3\arcsec $  ^{(c)}$  &79.1\% \\
{\scriptsize GALEX} & NUV, FUV & 5\arcsec $  ^{(e)}$ & 81.4\%\\
\hline
\end{tabular}\\
{(a)~\citet{Condon98}; (b) \citet{W97};
(c) \citet{Blanton05}; (d) respectively, following http://wise2.ipac.caltech.edu/docs/release/prelim/; (e) \citet{Budavari09}}; X-ray couterparts were not found in any of the ROSAT, Chandra or XMM catalogues (objects not observed or non-detected).
\end{table}

\noindent Spectral line related quantities were directly obtained from the MPA-JHU DR7 
release of spectral measurements\footnote{ http://www.mpa-garching.mpg.de/SDSS/DR7/}. All the retrieved flux measures have 
S/N$>$ 3, are corrected for stellar absorption 
\citep[following][]{Tremonti04}\,  and for foreground (galactic) reddening 
\citep[as in][]{ODonnell94}. We note that SDSS spectra were obtained with
 fibers 
that only cover the central 3 arcseconds of a galaxy. 
Considering that our objects are relatively nearby (z$<$0.06), that region is 
equivalent to a maximum of $\approx$ 3.4 kpc, therefore the spectral information for our objects is local, not an average measure for the whole galaxy.

\section{Red, Bulgeless and Active?}
Our main goal is to try to identify galaxies that overall have an old stellar population, no significant ongoing star formation, no prominent bulges, and show extra-activity in their centres (possibly capable of having dominated the quenching of the star formation). 
For that, we will next discuss the relevant quantities (redness, bulge significance, AGN content and degree of star formation) in terms of emission line diagnostics, colours
and structure.

\subsection{Diagnostic Diagrams}
\label{sec_diagnostic}

Diagnostic diagrams based on the emission line output are useful
to try to understand the main origin of the photoionisation: stellar or AGN
related.  The objects were classified based on the following diagrams:\\

\begin{figure}
\includegraphics[width=7.5cm]{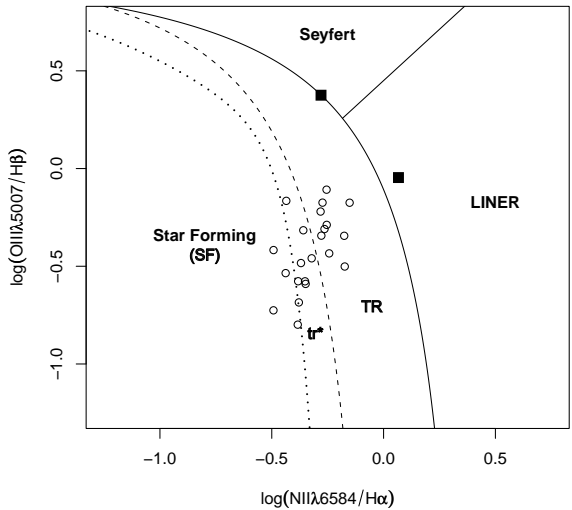}
\caption{
\label{fig_BPT1}BPT diagram representing 24 of the 43 sample objects (ie. those having these particular line fluxes with S/N$>$ 3). The solid curve represents the theoretical maximum starburst model from \citet{Kewley01}, the dashed curve represents the empirical limit between star forming galaxies and the ones with AGN activity proposed by \citet{Kauff03}. The galaxies between these curves can present simultaneously AGN and star formation activity: they are composite objects and henceforth labelled TR. That region can be enlarged if we consider the more conservative limit (dotted line) for pure star forming galaxies proposed by \citet{Stasi06}; objects lying between this line and the \citet{Kauff03} line are labelled tr*. The straight line represents the LINER/Seyfert separation limit proposed by \citet{Schawinski07}. The black squares are, according to this diagram, the best candidates to Seyfert/LINER in our sample.}
\end{figure}

\begin{figure}
\includegraphics[width=7.5cm]{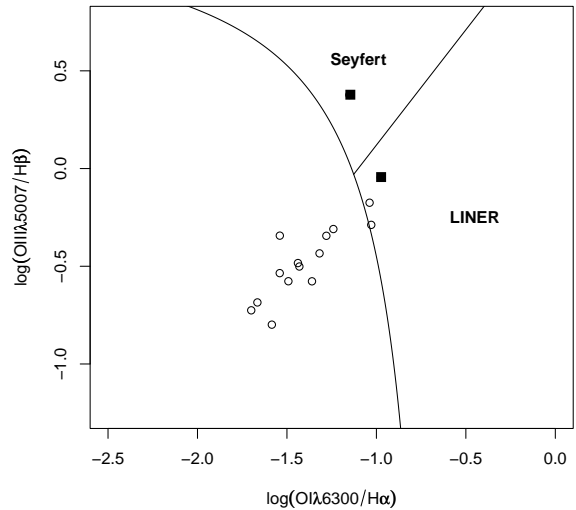}
\caption{\label{fig_BPT2}In this diagram there are 16 of the 43 sample galaxies. The solid curve separates star forming galaxies from AGN, according to \citet{Kewley01}. The straight line is the separation between Seyfert and LINER proposed by \citet{Kewley06}. The black squares are the best candidates to Seyfert/LINER according to the diagram in Figure \ref{fig_BPT1}.}
\end{figure}

\noindent {\bf BPT diagrams} \citep{BPT81}. We used the [NII]/H$_\alpha$ versus [OIII]/H$_\beta$ diagram, presented in 
Figure \ref{fig_BPT1}, that differentiates objects dominated by star formation emission from those dominated
by AGN emission. We also plotted the [OI]/H$_\alpha$  versus 
[OIII]/H$_\beta$  diagram, in Figure \ref{fig_BPT2}, that
separates ``standard'' AGNs from LINER \citep[eg.][]{Kewley06, Schawinski07}. Depending on their location in such diagrams, the objects are classified 
as: Star-forming galaxies (SF), composite objects (tr* and TR) {\em ie.} 
objects that may have both AGN and star formation activity, LINER and Seyfert-type (see figures \ref{fig_BPT1} and \ref{fig_BPT2} for details).

\begin{figure}
\includegraphics[width=7.5cm]{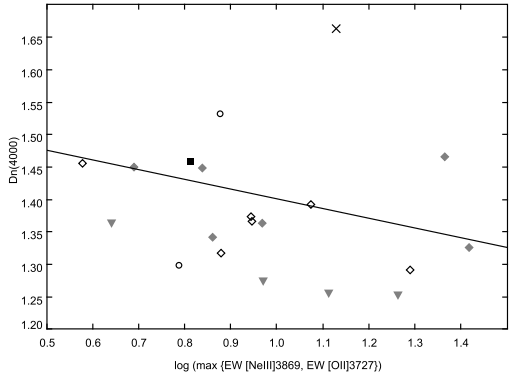}
\caption{\label{fig_DEWtest} Dn(4000) index as a function of the maximum of the equivalent widths of the [OII] and [NeIII] emission lines for the 37 objects in our sample that have these lines with S/N$\ge$3 measured flux. The line, given by equation (13) of \citet{Stasi06}, broadly separates AGN (above) from SF galaxies (below), according to the same authors. Symbols indicate the previous classification provided by the BPT diagrams, namely - black square: LINER; grey triangles: SF; cross: Seyfert; grey filled diamonds: TR; empty diamonds: tr*; open circles: unclassified objects. Based on this diagram, we further attributed the LINER classification to the previously unclassified object that stands out in the upper region of this plot.}
\end{figure}

\noindent {\bf DEW diagnostic} \citep{Stasi06}. In a complementary approach, objects were 
classified as star-forming galaxies or AGN hosts according to their location 
in a diagram of the 4000\AA\ break, Dn(4000)\footnote{computed as
the ratio of the mean flux at 4000-4100 \AA\ and 3850-3950 \AA\ bands following \citet{Balogh99} and directly available in the MPA-JHU catalogue.}, against the maximum of the equivalent widths of the [OII] or [NeIII] emission lines - see Figure \ref{fig_DEWtest}.\\

\begin{figure}
\includegraphics[width=8.5cm]{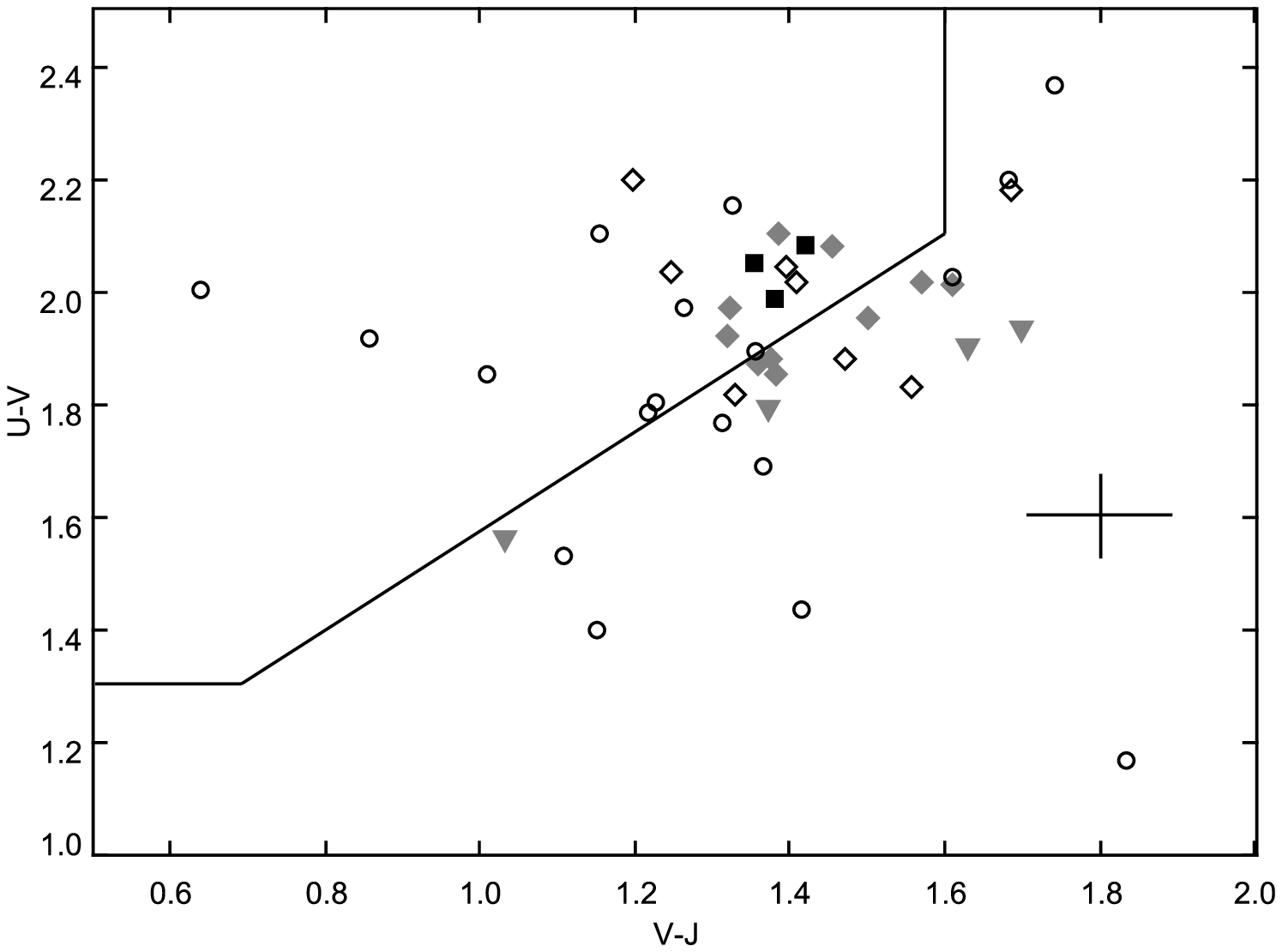}
\caption{\label{fig_Williams} Rest frame U-V versus V-J AB colours for our 43 objects. The lines indicate the separation between star-forming and quiescent galaxies according to \citet{Williams09}. The latter occupy the upper left part of this diagram. Symbols are - black squares: LINER + Seyfert; grey triangles: SF; grey filled diamonds: TR; empty diamonds: tr*; open circles: unclassified objects. On the right we indicate an average error bar.} 
\end{figure}

\noindent  Among the initial 43 objects, a total of 25 have detectable (ie S$/$N$>$3) 
emission lines relevant for the diagnostic diagrams and were classified in the categories listed above; the remaining ones will be referred to as "unclassified objects" 
hereafter. Considering all Figures - \ref{fig_BPT1}, \ref{fig_BPT2} and
\ref{fig_DEWtest} (see the captions for further details) -, we conclude that (1) the majority of the objects are composite {\em ie.} these are objects with relatively weak emission lines, and in which AGN-related 
photoionisation and stellar-related photoionisation might co-habit; 
(2) only a few objects -- 4 -- are classified as pure star-forming galaxies; (3)
in 2 objects there is a LINER nuclei and in 1 a Seyfert-like nucleus,
indicating the presence of extra activity in their centres (see discussion in 
section \ref{agns}). All final classifications resulting from BPT and DEW are given in table \ref{table_sample}. 

\begin{figure*}
\includegraphics[height=4cm,width=4.cm]{J075117_oplot.epsi}
\includegraphics[height=4cm,width=4.cm]{J080217_oplot.epsi}
\includegraphics[height=4cm,width=4.cm]{J093159_oplot.epsi}
\includegraphics[height=4cm,width=4.cm]{J095146_oplot.epsi}
\caption{\label{imagespluscontours} r-band SDSS images with the FIRST-VLA radio contours 
superimposed, with a range of 10 levels between the maximum and minimum intensity (1.0$\times10^{-4}$Jy), for,
from the left to the right: at z$\sim$ 0.06 SDSSJ075117.08+324425.1 (LINER), SDSSJ080217.94+112535.0 (SF), at z$\sim$ 0.03
SDSSJ093159.95+512254.0 (LINER) and SDSSJ095146.53+273245.8 (SF)}
\end{figure*}
  
\subsection{On the red nature of the colours}\label{secdust}
\label{sec_redness}

Attesting the robustness of our sample includes verifying the red nature of 
the  galaxies - whether it is due to older stellar populations or rather related with dust extinction - since, despite the 
selection criteria, the sample does contain some 
star-forming objects according to the previous diagnostic diagrams.
We  investigate the origin of the red
colours following \citet{Williams09}, by using their proposed U-V versus V-J  
diagram  to distinguish between 
dusty star-forming and red quiescent galaxies. The results
are presented in Figure \ref{fig_Williams},
 where we applied the transformation equations given in \citet{BlantonRoweis07} 
to convert from u and g to U and V, respectively; J was directly assessed from 2MASS.

\noindent The  objects are relatively 
concentrated at the interface between the star-forming region and the quiescent region and, consistently, SF objects only occupy the ``star-forming region''. The ``quiescent region'' of the diagram is populated by both LINER and 
Seyfert objects, plus some of the composite and unclassified objects - there are 17 objects in these conditions (we do not consider the 3 unclassified objects that lie on top of the line). We will henceforth consider this subsample 
of 17 objects as our best candidates of 
SF-quenched, massive and bulgeless galaxies.

\subsection{AGN candidates}
\label{agns}

 \noindent Concerning
the search for extra activity in the centre of these galaxies, and besides
the optical emission line diagnostic discussed in section \ref{sec_diagnostic},
 the 
radio band offers one of the best complementary probes for inferring 
the possible  presence of extra activity. A compact core radio morphology is 
a good indicator of the presence  of  an AGN, and this has been detected in very faint objects \citep[eg.][]{Nagar05,Falcke00}, whereas an extended radio 
morphology  would rather argue in favour of a starburst.  Among the group
of 17 quiescent galaxies (from section \ref{sec_redness}), 6 objects show compact radio morphology
in FIRST VLA maps (5$\arcsec$ resolution):  4 are classified as composite, 
and 2 as LINER, ie, all have low ionization emission lines. The following discussion concerns those 6 radio emitters plus the Seyfert-like object. The nature
of the low ionization emission lines is not a settled issue, and for example 
the recent work of \cite{Yan2012} argues that the photoionisation
 might be partially stellar in origin. 
In our case, the fact that the objects show compact radio morphology
and were classified as quiescent in the optical-NIR colour plane argues for at least the
co-existence of an AGN. Their compact radio morphology  contrasts with
the rather extended one that is observable for the SF galaxies of our sample, as illustrated in Figure
 \ref{imagespluscontours}, where we present the r-band SDSS images with the
 FIRST VLA contours superimposed for 
 2 pairs of LINER and SF objects lying at similar redshifts for a direct 
comparison.\\
We next further characterise these 7 candidate bulgeless, SF-quiescent galaxies with an AGN.


\begin{figure*}
\center{
\includegraphics[height=4cm]{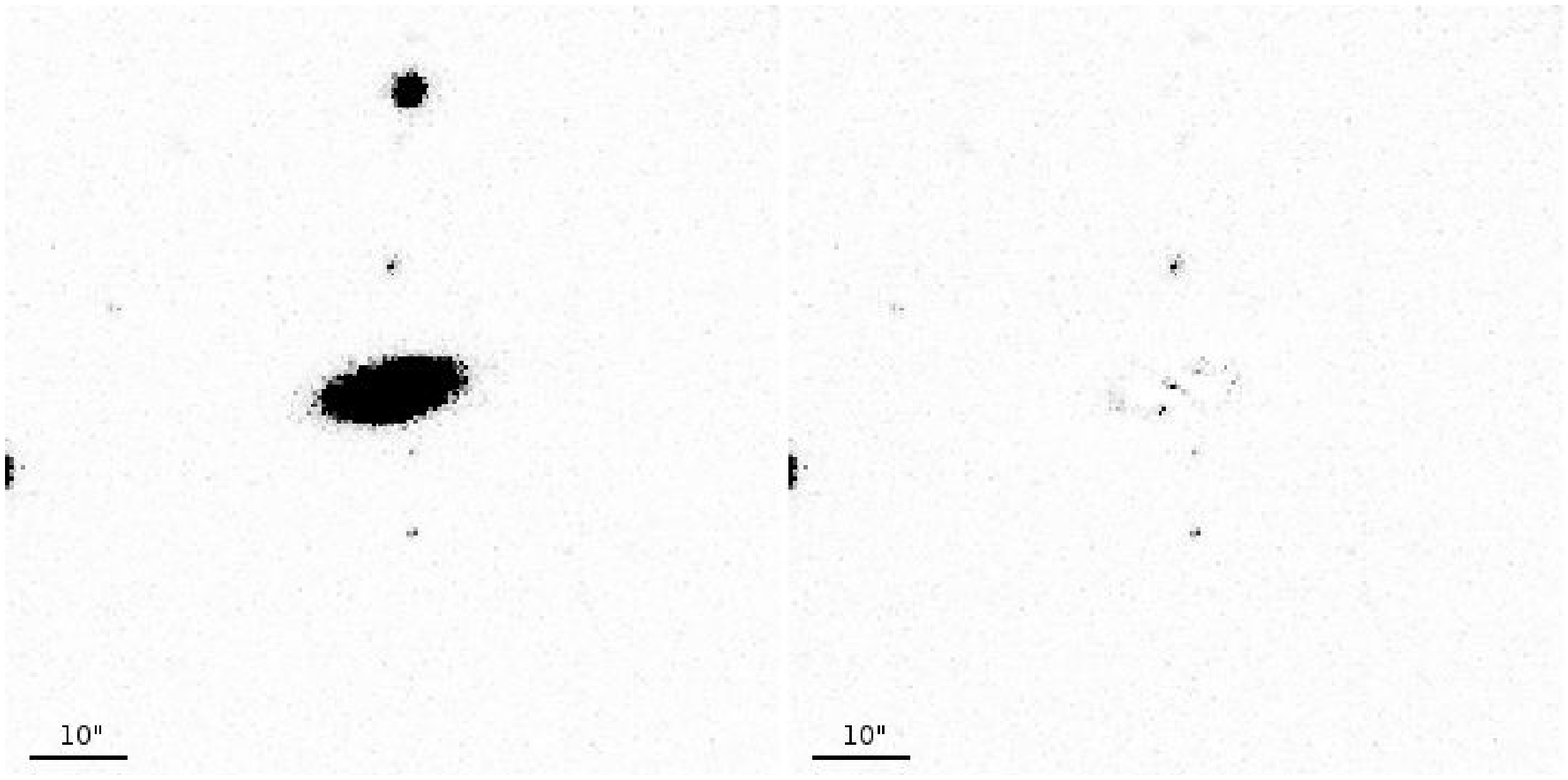}
\includegraphics[height=4cm]{galaxy_31_2comps.epsi}\\
\includegraphics[height=4cm]{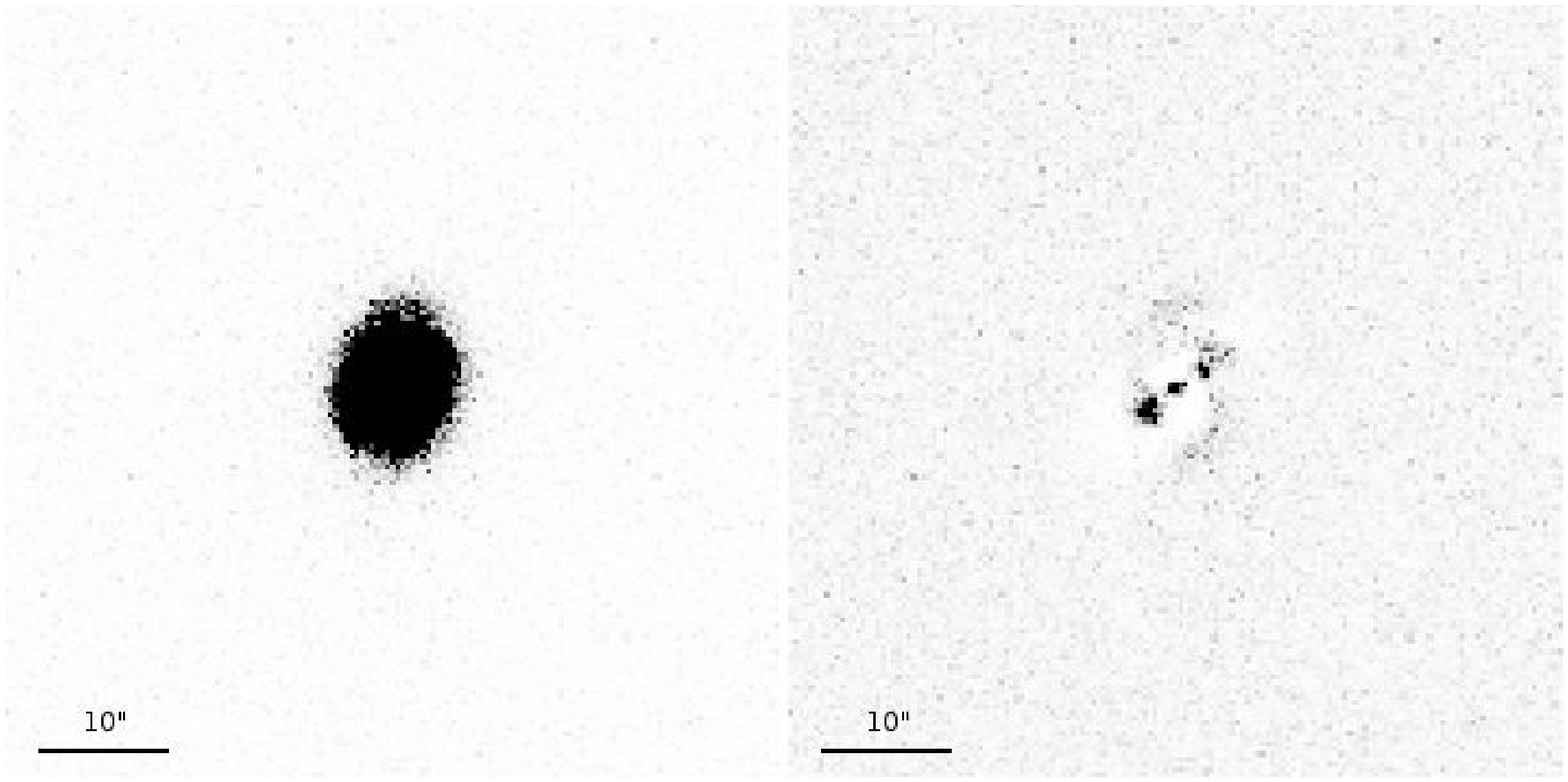}
\includegraphics[height=4cm]{galaxy_6_disk.epsi}\\
\includegraphics[height=4cm]{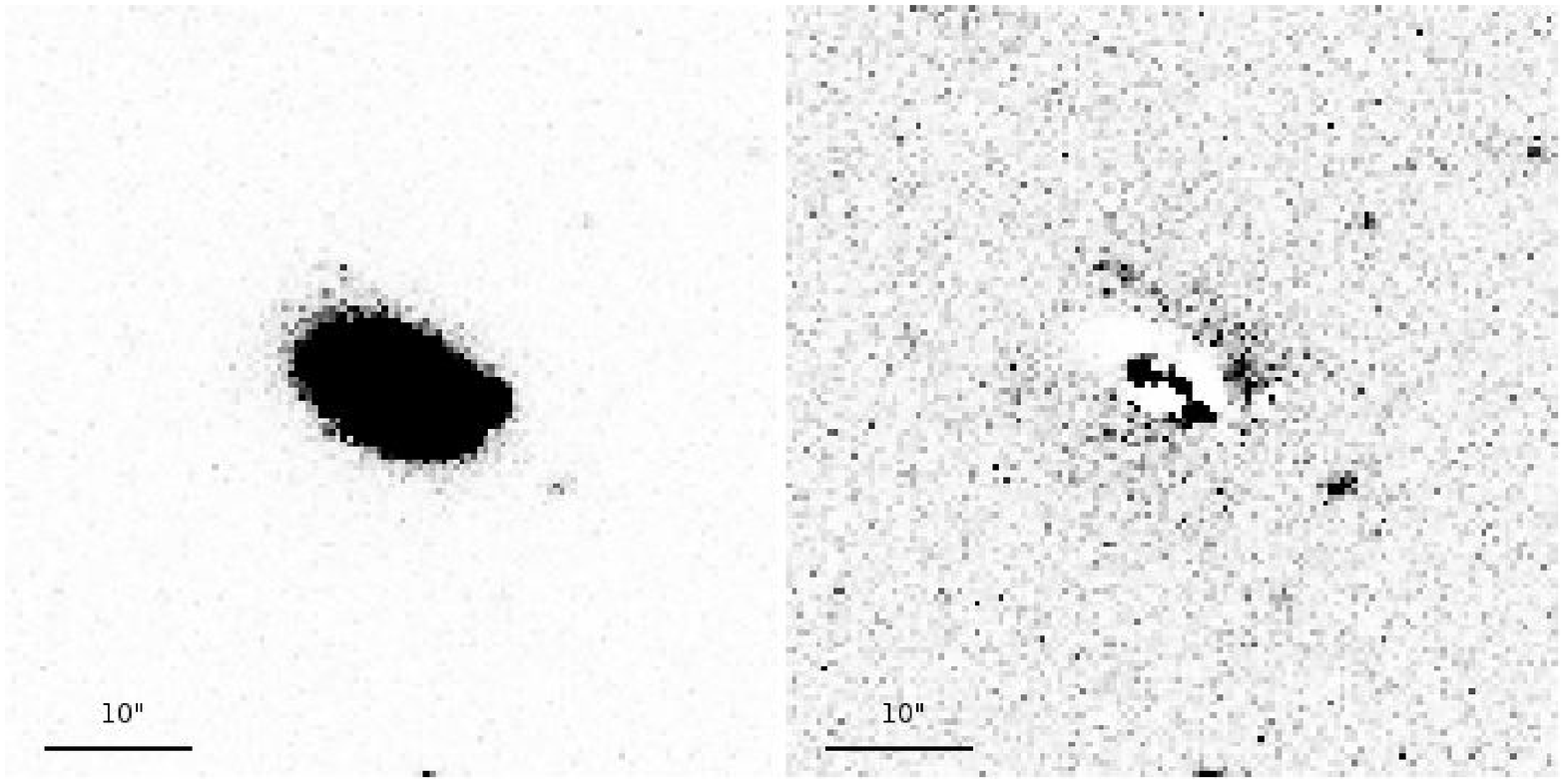}
\includegraphics[height=4cm]{galaxy_53_disk.epsi}\\
\includegraphics[height=4cm]{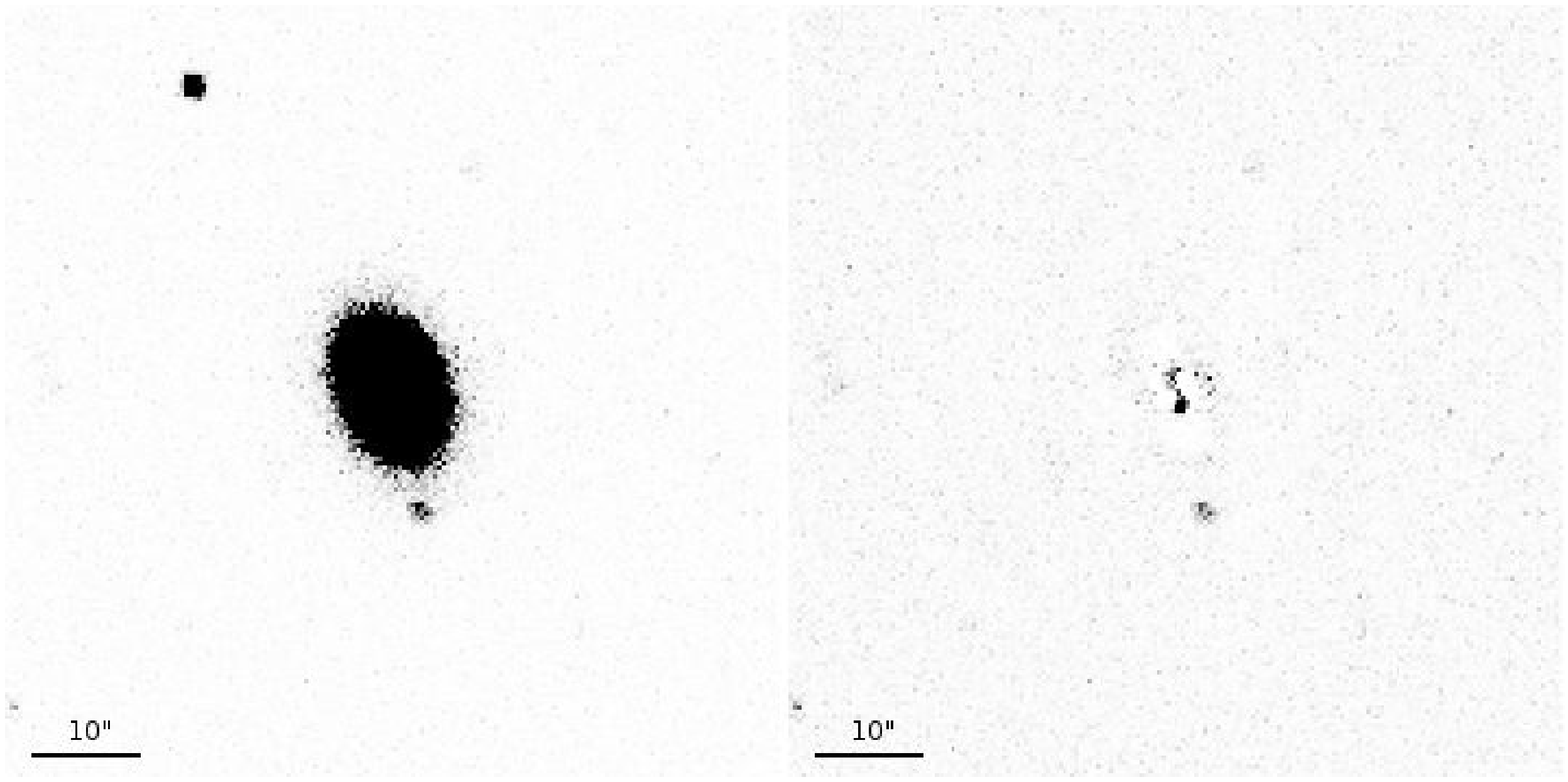}
\includegraphics[height=4cm]{galaxy_18_2comps.epsi}\\
\includegraphics[height=4cm]{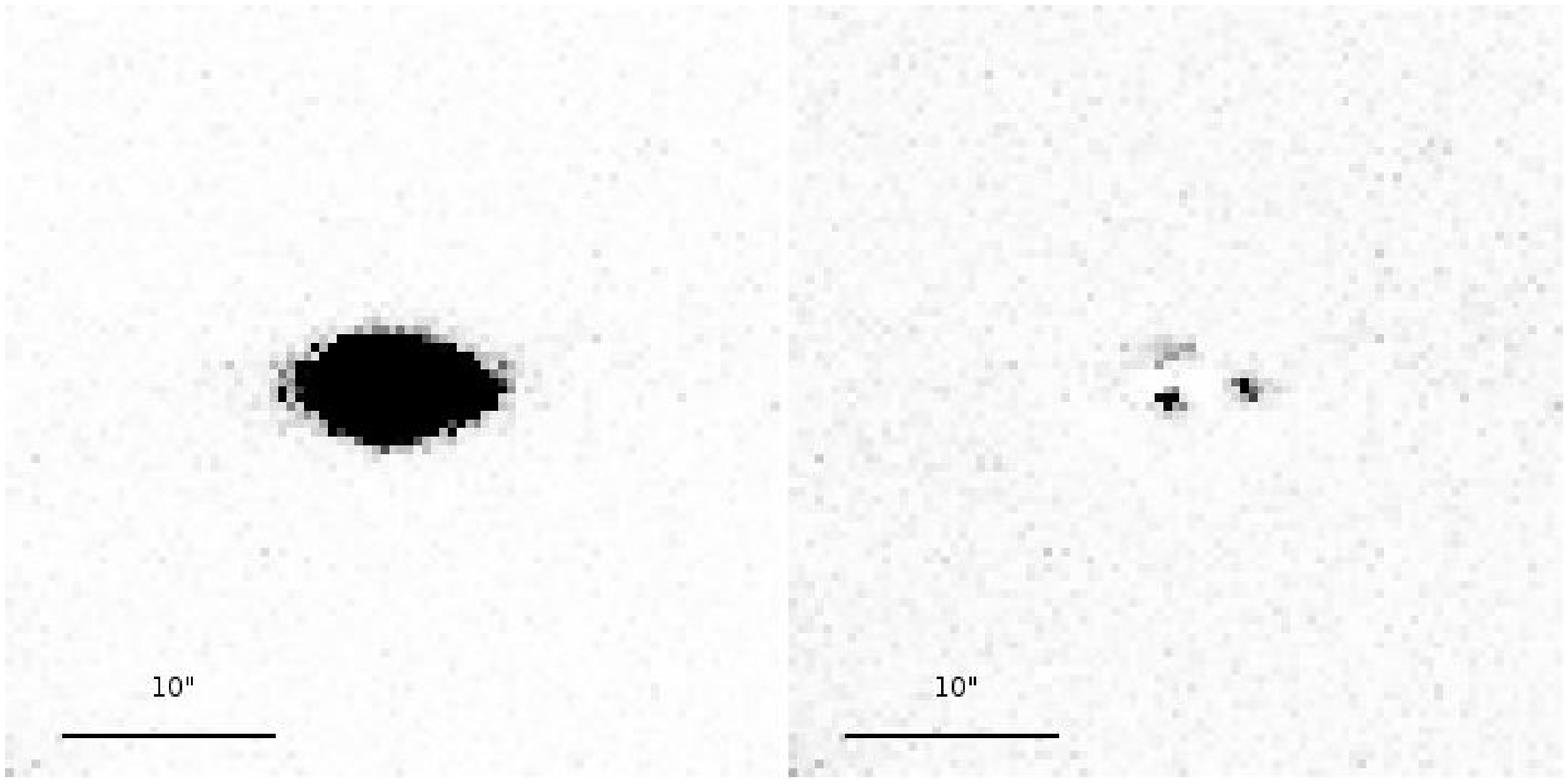}
\includegraphics[height=4cm]{galaxy_50_disk.epsi}\\
}
\caption{\label{sed_eg1} --
{\bf Left- }SDSS r-band images; {\bf Middle- } the residual images  
obtained after subtracting the best-fit model in each case (see table \ref{tab-sersic}), displayed in the same scaling as the original image –- residuals are always below 2\% and we note that often, these do not solely come from a bulge but 
also from structure in general (eg. spiral arms) that we deliberately did not attempt to model;  
{\bf Right- }
the fitted 2D profile, where circles represent
the data, full line the total fitted profile (separated in 2 components only in the case of 2 galaxies),  
for the following 
objects, from {\bf Top to Bottom}: 
SDSS J075117.08+324425.1 (LINER), SDSS J080441.34+454715.6 (TR),
SDSS J082919.82+061744.8 (tr*), SDSS J093159.95+512254.0 (LINER),
SDSS J094208.40+094355.5 (Seyfert)}
\end{figure*}

\begin{figure*}
\center{
\includegraphics[height=4cm]{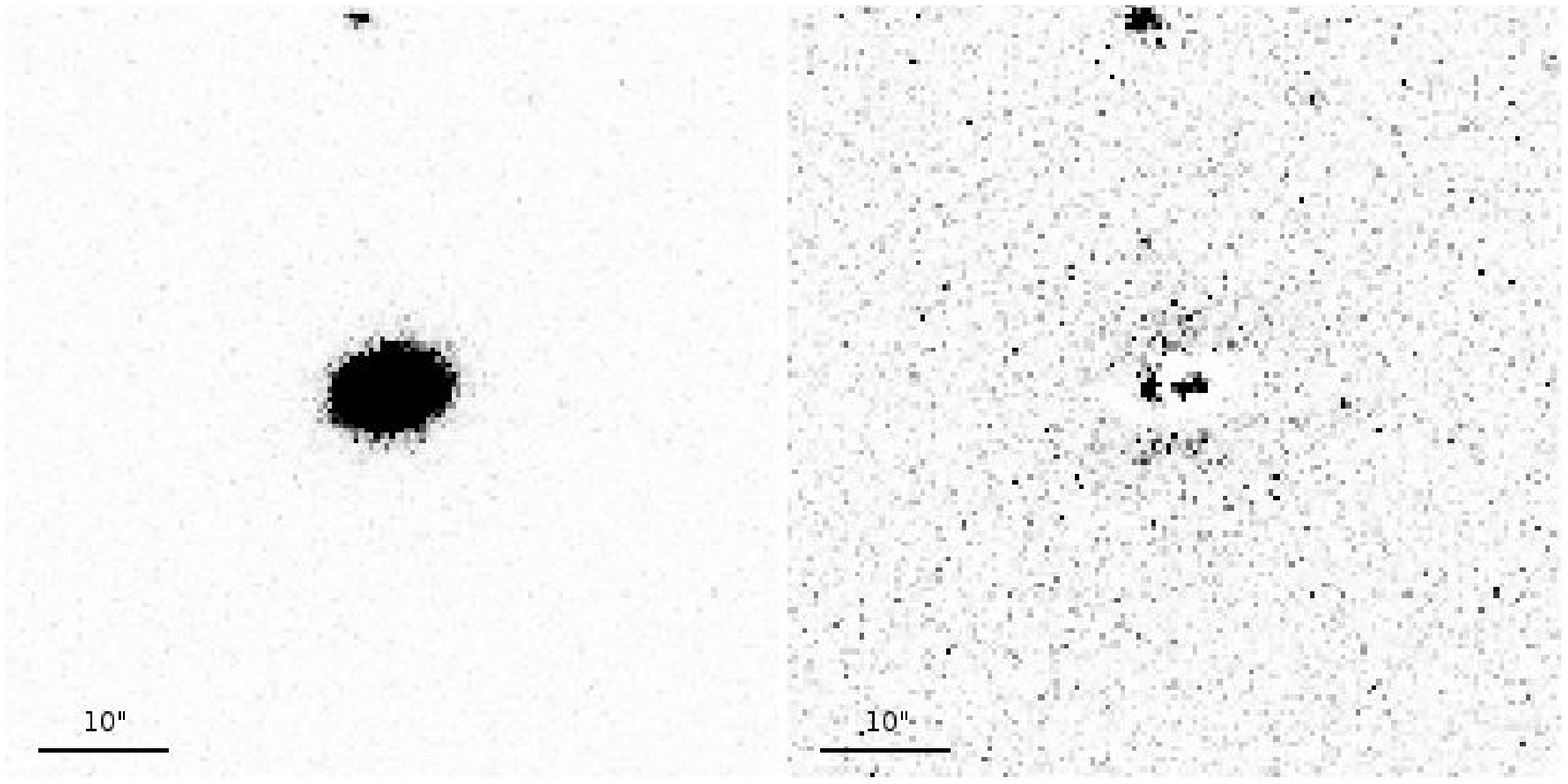}
\includegraphics[height=4cm]{galaxy_54_disk.epsi}\\
\includegraphics[height=4cm]{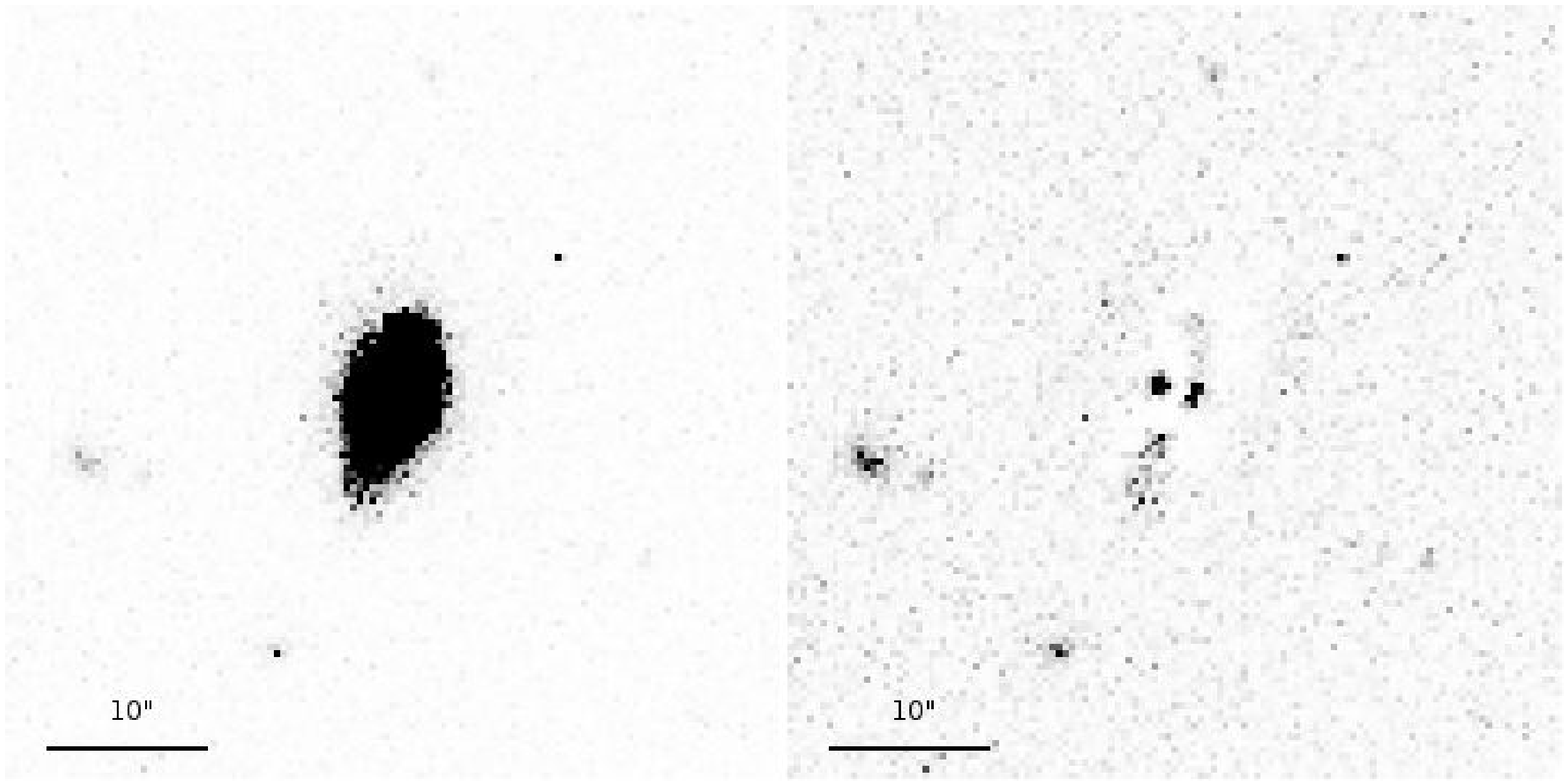}
\includegraphics[height=4cm]{galaxy_1_disk.epsi}
}
\caption{\label{sed_eg2} --
as in Figure \ref{sed_eg1} for the following 
objects, from {\bf Top to Bottom}: 
 SDSS J140929.47+000837.2 (TR) 
and SDSS J103422.30+442349.1 (tr*)}
\end{figure*}

\subsubsection{Bulgeless nature}
 For this selected group of 7 objects we re-assessed their morphological
structure by performing a detailed 2D analysis of their r-band images 
 with GALFIT \citep{Peng2010} using PSF convolution. A disk
profile (S\'ersic law with n=1) was assumed for all 7 objects, and 
the residuals, if existent, were modelled with an additional component (S\'ersic law with free n). The results are presented in Figures \ref{sed_eg1}, \ref{sed_eg2},
and in Table \ref{tab-sersic}. All but 2 objects are well described by a 
single disk component; in those two cases, there could also be some sort of bulge-like structure where the corresponding S\'ersic index is far below n$=$1.5.

To verify the robustness of the fits, we recovered K-band data from UKIDSS \citep{Lawrence07} for 3 of these 7 galaxies (those having these data available) and ran GALFIT on those images. The output parameters being totally consistent with the ones derived from the r-band SDSS images, gives us some confidence that the above overall results are robust, so we proceeded with the analyses provided using the r-band derived models.

The next step consisted in quantifying the excess light in the central part of each galaxy relatively to a single disk model - thus giving an upper limit for any putative bulge-like component. This was done in a manner that is consistent with \citet{FisherDrory10}, i.e. we (i) masked the central region of the galaxy in order to fit a disk model to its outer regions, 
(ii) extrapolated this fit to the masked inner part to obtain the contribution 
of the disk to the light in the "bulge" (or central) region, (iii) subtracted this contribution to the total light of the galaxy limited to the same central region, thus obtaining the light that is in excess (relatively to a single disk) in this central region. We then computed the ratio between this excess light to the total light of the galaxy – results are given in Table \ref{tab-residuals} and confirm that 6 out of our 7 galaxies remain well described by a single exponential disk. For these 6, the excess (central) light remaining after subtraction of the disk always accounts from 2\% to less than 5.3\% of the total galaxy light, which is totally consistent with the results of eg. \citet{Simmons2013}. Again, we stress that this estimate is an upper limit to the light coming from a possibly existing bulge-like component, since at least part can come from any type of structure in the centre. We further note that in the 6 cases, only one galaxy (SDSS J075117.08+324425.1, the one with the highest percentage of light excess, 5.3\%) accomodates  an additional S\'ersic model with low n (see Table \ref{tab-sersic}). The 7th galaxy (SDSS J093159.95+512254.0) indeed seems to host central low surface brightness structure (S\'ersic component with n=1.08 – Table \ref{tab-sersic}); residuals for this galaxy indicate 25\% of light in excess relatively to a pure disk -- Table \ref{tab-residuals}, so one can consider revising its classification from bulgeless into “pseudo-bulge galaxy”. There is, however, no sharp quantitative frontier between the two classes - that share the absence of a classical bulge and are clearly dominated by a disk \citep[and are generally classified as bulgeless by other authors, eg.][]{Sachdeva13} -, so we will retain all 7 objects for the following analysis.

Finally, and still considering the central light, as estimated above, if we wish to check how much of it is accounted for by a point-like source (putative AGN), then models indicate that this would contribute with no more than 1\% to the total galaxy light \citep[as expected, see eg.][]{DroryFisher07,Phillips96}.

\subsubsection{Level of star formation}
We can try to attest the significance of the ongoing SFR by looking at the 
NUV-r colour index  (as a function of stellar mass), following 
\citet{Cortese12}, and by estimating a star formation rate. This author
 showed the importance of assessing the current star formation through 
indicators relying on data (eg. NUV, MIR) other than just in the optical waveband. Five among the 7 objects have NUV data from GALEX and 
their colours locate them in the green valley region ($3.5 \la$ NUV-r $\la 5$), 
as likely expected for disk galaxies that no longer have significant ongoing star 
formation -- we further note that composite AGN-hosting galaxies show such 
colours \citep[eg.][]{Martin07}. Consistently, their SFR -- estimated using 
$22 \mu$m WISE data and equation (10) of \citet{Rieke09} -- are low even assuming, as an upper limit, that all the MIR/$22 \mu$m flux is due to star formation: less 
than 1 M$\sun$/yr for 5 and marginally higher for the other 2 galaxies 
(1.8 and 2.2 M$\sun$/yr).

\subsubsection{On the SF-quenching}

\noindent As discussed in section \ref{sec_redness},
the objects were selected as quiescent-type according
to their location in the Williams et al. colour-colour diagram. We verified that, consistently, their NUV-r colours place them in the green valley and they have no significant recent or ongoing star formation activity (see section \ref{agns}). Could this quenching be the consequence of AGN-feedback?
\citet[][see also Bower et al. 2006]{Croton06} introduced a type of AGN-feedback,  
the ``radio mode feedback'', that they 
argue is most efficient when the virial velocity of the halo
 v$_{\mbox{vir}}$ is  $>$ 100 kms$^{-1}$  at z$\sim$0 (see their Figure 7).
To verify whether our galaxies could be accommodated, in a reasonable way, in a 
  scenario of [mild] AGN feedback, we have made an estimate of their v$_{\mbox{vir}}$, assuming 
M$_{\mbox{vir}}$ $\sim$ 10 M$_\ast$ and equation (2) of \citet{Croton06},
and obtained values between 72 kms$^{-1}$ and 96 kms$^{-1}$.
These values  are marginally compatible with efficient triggering of radio  
mode AGN feedback \citep{Croton06} so, given the adopted assumptions, we think that it is plausible  
that AGN activity, possibly even more significant in the past relatively to the ongoing measured level,  
may have affected the star formation efficiency.
Just to have an indication, we have computed 
rough black hole 
mass limits for our bulgeless AGN candidates with Seyfert and LINER spectra 
as follows: we estimated an upper limit for the AGN luminosity by 
the [OIII] emission line luminosity and by considering the bolometric correction $L_{\it{bol}}/L_{[OIII]}\sim$ 600 
\citep[valid for LINER and Seyfert,][]{KauffHeck09}. 
Then a black hole mass 
was calculated by assuming that the objects are radiating at the Eddington limit (which is an extreme
case). Only two of the three objects have the required data - the estimates 
are presented in Table \ref{table_sample} and indicate small BH masses.
Finally, we note that in satellite galaxies the star formation 
quenching may be dominated by mechanisms other than AGN-related \citep[such as stripping of gas or other, see e.g.][]{Cole00}. Among
our objects, only two are not central in their dark matter halos,  
SDSS J094208.40+094355.5 (the Seyfert) and  
SDSS J103422.30+442349.1 (a tr* object), according to the classification 
in the DR7 group catalogue (X. Yang, private communication). Whereas for the tr* object this could cast doubt on the AGN being the dominant feedback mechanism ruling the star formation quenching, the Seyfert galaxy, having definite signature of hosting an AGN, remains as a strong candidate for this scenario. We thus have a very small but rather robust set of galaxies in which, likely, a SMBH hosted by an insignificant or absent bulge may have played a major role in quenching their activity of forming stars.

\begin{table}
\caption{Structural parameters issued from the 2-D S\'ersic profile fits to the 7 objects presented in Figures \ref{sed_eg1} and \ref{sed_eg2}: (1) 
effective radius, in arcsecs, of the  disk model corresponding to a S\'ersic law 
with n=1; for two galaxies a second component was needed and for it we 
present (2) the respective S\'ersic index and (3)
the effective radius, in arcsecs, of that second component. (4) gives the  $\chi^{2}/\mu$ of
the global fit.}
\label{tab-sersic}
\centering
\begin{tabular}{ccccc} 
\hline
SDSS name & $R_{e,d}$ & $n_{b}$ & 
$R_{e,b}$ & $\chi^{2}/\mu$ \\
 & (1) & (2) & (3) & (4) \\
\hline
SDSS J075117.08+324425.1 & 4.8 & 0.27 & 1.4 & 1.16 \\ 
SDSS J080441.34+454715.6 & 3.8 & & & 1.09 \\ 
SDSS J082919.82+061744.8 & 3.2 & & & 1.25 \\ 
SDSS J093159.95+512254.0 & 5.9 & 1.08 & 1.9 & 1.02 \\ 
SDSS J094208.40+094355.5 & 2.9 & & & 1.10 \\ 
SDSS J103422.30+442349.1 & 2.1 & & & 1.10 \\
SDSS J140929.47+000837.2 & 3.5 & & & 1.02 \\ 
\hline
\end{tabular}
\end{table}

\begin{table}
\caption{Ratio of the central residuals to the total light; residuals here are those obtained after subtraction of a pure disk in the central part of the galaxy, in a similar fashion as done by \citet{FisherDrory10} -– see text.}
\label{tab-residuals}
\centering
\begin{tabular}{cc} 
\hline
SDSS name & (central residual)/total light in \% \\
\hline
SDSS J075117.08+324425.1 & 5.31\% \\ 
SDSS J080441.34+454715.6 & 1.34\% \\ 
SDSS J082919.82+061744.8 & 1.61\% \\ 
SDSS J093159.95+512254.0 & 25.43\% \\ 
SDSS J094208.40+094355.5 & 1.70\% \\ 
SDSS J103422.30+442349.1 & 0.63\% \\
SDSS J140929.47+000837.2 & 0.81\% \\ 
\hline
\end{tabular}
\end{table}

\section{Summary and Conclusions}
In this work we present the properties of a sample of red and bulgeless 
massive galaxies. Through a very conservative approach we selected 43 
such galaxies from the NYU-VAGC catalogue issued for the SDSS DR7. 
The origin of the redness of their colours was discussed in terms
of the location of the objects in the UV-VJ diagram, and we identify a group 
located in the quoted quiescent zone, 17 objects in total. 
These represent the best candidates
for red (as for star-forming quenched) and bulgeless massive galaxies. 
The majority   show emission lines such that they are classified as composite, 
LINER and Seyfert.  Six out of these 17 objects are radio emitters that 
show  compact radio morphology in FIRST-VLA maps. Even if FIRST resolution 
does not permit to rule out the existence of a very central star-forming 
region in these objects, the fact is that their compact emission is very 
different from the extended radio emission shown by the SF objects of the 
sample lying at  equivalent redshifts.  We thus consider the 6 radio compact 
emitters plus the object classfied as Seyfert by BPT diagrams as our best AGN hosting candidates; their current SFRs, as estimated considering that their MIR/$22 \mu$m  flux is thermal and SF-related, are low as well.  For this subsample of 7 galaxies we re-confirmed their
bulgeless nature (or pseudo-bulge in one case) by performing a dedicated 2-D profile fitting. 
Crude black hole mass estimated limits for a couple of the objects indicate 
M$_{\tiny\mbox{BH}}$$<$10$^{4.5}$ M$_\odot$).\\
\noindent In conclusion, 
following a very conservative approach, we identify 7 red bulgeless massive AGN-hosting candidates, 3 of them being classified by diagnostic diagrams as 
having LINER/Seyfert -type activity in their centre, the remaining 4 being composite objects, i.e. in which an AGN might co-exist. Different indicators point
to all of them having low SFRs.  AGN-feedback phenomena might
be the main responsible for the SF-quenching in 5 of them. The remaining two are satellite galaxies, and for those other mechanisms might contribute as well to the
 SF-quenching.

\section*{Acknowledgments}

The authors acknowledge technical support from the EURO-VO AIDA project in which framework this work was initiated and, 
in particular, E. Hatziminaoglou, M. Taylor and the ASTROGRID VO-Desktop helpdesk. M. Blanton is acknowledged for support concerning the NYU-VAGC usage. It is also a pleasure to acknowledge J. Brinchmann for helpful information on the SDSS data and MPA-JHU catalogues, and X. Yang for kindly giving us access to the DR7 group catalogs. We also thank the referee for a critical and careful reading that allowed improving the paper. The authors acknowledge 
financial support from project PTDC/CTE-AST/105287/2008 from FCT, Portugal. S. Ant\'on acknowledges the support from FCT through the projects PEst-OE/CTE/UI0190/2011 and Ciencia2007.
B.C. was partially supported by grant CAUP-07/2008-BII.\\

Funding for the SDSS and SDSS-II has been provided by the Alfred P. Sloan Foundation, the Participating Institutions, the National Science Foundation, the U.S. Department of Energy, the National Aeronautics and Space Administration, the Japanese Monbukagakusho, the Max Planck Society, and the Higher Education Funding Council for England. The SDSS Web Site is http://www.sdss.org/.\\
The SDSS is managed by the Astrophysical Research Consortium for the Participating Institutions. The Participating Institutions are the American Museum of Natural History, Astrophysical Institute Potsdam, University of Basel, University of Cambridge, Case Western Reserve University, University of Chicago, Drexel University, Fermilab, the Institute for Advanced Study, the Japan Participation Group, Johns Hopkins University, the Joint Institute for Nuclear Astrophysics, the Kavli Institute for Particle Astrophysics and Cosmology, the Korean Scientist Group, the Chinese Academy of Sciences (LAMOST), Los Alamos National Laboratory, the Max-Planck-Institute for Astronomy (MPIA), the Max-Planck-Institute for Astrophysics (MPA), New Mexico State University, Ohio State University, University of Pittsburgh, University of Portsmouth, Princeton University, the United States Naval Observatory, and the University of Washington.\\
Along the process described below we used various VO inter-connect
tools and applications, in particular the ASTROGRID VO-Desktop, TOPCAT and STILTS, that enable to
mine and analyse a large amount of tabular data.\\
This publication makes use of data products from the Wide-field Infrared Survey Explorer, which is a joint project of the University of California, Los Angeles, and the Jet Propulsion Laboratory/California Institute of Technology, funded by the National Aeronautics and Space Administration. \\
This research has made use of the NASA/IPAC Extragalactic Database (NED) which is operated by the Jet Propulsion Laboratory, California Institute of Technology, under contract with the National Aeronautics and Space Administration.

\input{table1.tex}

%

\label{lastpage}
\end{document}

%% file: table1.tex
\begin{table*}
\renewcommand{\tabcolsep}{0.25cm}
\caption[]{The sample of the initial 43 candidates obtained by applying the  
selection described in section 2. : (1) SDSS name (2), (3) ra, dec coordinates in degrees J 2000.0 (4) Colour index (5) Redshift (6) S\'ersic index (r-band, Blanton et al. 2005) (7) Inclination parameter (8) Petrosian absolute r-band magnitude (9) Logarithm of the mass of the stellar component measured in solar masses (10) BPT or DEW classification 
(11) estimate of the Black Hole mass assuming $L_{bol}/L_{[OIII]}\sim 600$ (Kauffmann \& Heckman, 2009), in M$_{\sun}$ units (12) Black hole mass uncertainty (\%) }
\label{table_sample}
\tiny
\begin{tabular}{lrrrrrrrrlrr}
\hline 
\scriptsize{Name} & ra & dec & g-r &  z&  n& q$_{am}$  & \scriptsize{MAG$_r$}&  \scriptsize{$log M_*$} &\scriptsize{BPT/DEW} &
\scriptsize{$log M_{BH}$} & $\Delta$(\scriptsize{$log M_{BH}$)} \\
(1) &(2)  &(3) & (4) & (5) & (6)  & (7) & (8) & (9) & (10) & (11) & (12)  \\
\hline
\scriptsize{SDSS-J003018.19-003008.1} &    7.5758   &    -0.5022 &    0.705 &    0.059 &     1.388 &     0.591 &     -19.859 &     10.18  &     -       
&  - &  -  \\
\scriptsize{SDSS-J010303.55+132950.3} &    15.7648  &    13.4973 &    0.71  &    0.058 &     1.44  &     0.63  &     -19.466 &     10.028 &     -       
&  -  &  -  \\
\scriptsize{SDSS-J011500.27+000151.3} &    18.7511  &    0.0309  &    0.758 &    0.05  &     1.462 &     0.844 &     -19.912 &     10.258 &     -       
&  -   &   -   \\
\scriptsize{SDSS-J011834.14-001341.7} &    19.6422  &    -0.2282 &    0.782 &    0.047 &     1.247 &     0.557 &     -20.341 &     10.457 &     TR      
&  -   &  -  \\
\scriptsize{SDSS-J033021.75+001547.1} &    52.5906  &    0.2630   &    1.163 &    0.037 &     1.224 &     0.694 &     -19.865 &     10.684 &     -      
&   -   &  -  \\
\scriptsize{SDSS-J072403.09+404833.5} &    111.0128 &    40.8093 &    0.965 &    0.05  &     1.076 &     0.822 &     -19.295 &     10.239 &     SF      
&  -    &   -   \\
\scriptsize{SDSS-J074600.04+214323.2} &    116.5001 &    21.7230  &    1.372 &    0.046 &     1.314 &     0.58  &     -18.059 &     10.19  &     -      
&   -   &   -   \\
\scriptsize{SDSS-J075117.08+324425.1} &    117.8211 &    32.7403 &    0.764 &    0.056 &     1.14  &     0.585 &     -20.152 &     10.361 &     LINER   
&  -    &   -   \\
\scriptsize{SDSS-J080217.94+112535.0} &    120.5747 &    11.4264 &    0.838 &    0.06  &     1.151 &     0.936 &     -19.555 &     10.203 &     SF      
&  -    &   -   \\
\scriptsize{SDSS-J080441.34+454715.6} &    121.1722 &    45.7876 &    0.733 &    0.05  &     1.331 &     0.729 &     -20.351 &     10.407 &     TR      
& -  &  - \\
\scriptsize{SDSS-J082919.82+061744.8} &    127.3326 &    6.2957  &    0.751 &    0.048 &     1.303 &     0.574 &     -20.419 &     10.453 &     tr*     
&    -  &   - \\
\scriptsize{SDSS-J083639.67+471515.3} &    129.1652 &    47.2542 &    0.846 &    0.053 &     1.337 &     0.535 &     -19.129 &     10.042 &     tr*     
&   -   &    -   \\
\scriptsize{SDSS-J084251.31+525530.0} &    130.7137 &    52.9250  &    0.81  &    0.059 &     1.418 &     0.632 &     -20.166 &     10.418 &     -      
&   -   &    -   \\
\scriptsize{SDSS-J085640.72+055235.4} &    134.1696 &    5.8764  &    0.758 &    0.059 &     1.233 &     0.578 &     -19.366 &     10.04  &     tr*     
&   -   &    -   \\
\scriptsize{SDSS-J093159.95+512254.0} &    142.9998 &    51.3817 &    0.699 &    0.033 &     1.498 &     0.823 &     -19.62  &     10.077 &     LINER   
&   3.378  &  24.1\% \\
\scriptsize{SDSS-J094208.40+094355.5} &    145.5350  &    9.7320   &    0.789 &    0.059 &     1.409 &     0.546 &     -19.615 &     10.173 &     Seyfert
&  4.458   &   8\% \\
\scriptsize{SDSS-J095146.53+273245.8} &    147.9439 &    27.5460  &    0.75  &    0.033 &     1.072 &     0.855 &     -19.845 &     10.223 &     SF     
&   -    &   -   \\
\scriptsize{SDSS-J095517.41+174114.7} &    148.8225 &    17.6874 &    0.802 &    0.045 &     1.387 &     0.546 &     -19.815 &     10.268 &     SF      
&   -    &   -   \\
\scriptsize{SDSS-J101422.68+182650.6} &    153.5945 &    18.4473 &    0.883 &    0.044 &     1.439 &     0.626 &     -19.054 &     10.053 &     tr*     
&  -     &   -    \\
\scriptsize{SDSS-J103422.30+442349.1} &    158.5929 &    44.3970  &    0.723 &    0.052 &     1.4   &     0.665 &     -19.808 &     10.178 &     tr*    
&  -& - \\
\scriptsize{SDSS-J103543.35+121518.1} &    158.9306 &    12.2550  &    0.886 &    0.05  &     1.387 &     0.57  &     -20.074 &     10.463 &     -      
&   -    &    -   \\
\scriptsize{SDSS-J103856.94+254521.9} &    159.7372 &    25.7560  &    1.141 &    0.051 &     1.057 &     0.821 &     -19.206 &     10.395 &     -      
&   -    &   -    \\
\scriptsize{SDSS-J103957.42+174019.5} &    159.9892 &    17.6720  &    0.697 &    0.057 &     1.197 &     0.62  &     -19.589 &     10.062 &     -      
&   -    &   -    \\
\scriptsize{SDSS-J110313.24+074253.7} &    165.8051 &    7.7149  &    0.729 &    0.055 &     1.473 &     0.819 &     -19.919 &     10.23  &     -       
&   -    &   -    \\
\scriptsize{SDSS-J110635.18+440248.7} &    166.6465 &    44.0468 &    0.746 &    0.037 &     1.425 &     0.543 &     -19.338 &     10.015 &     tr*     
&   -    &   -    \\
\scriptsize{SDSS-J115759.73+250931.4} &    179.4989 &    25.1587 &    0.742 &    0.058 &     1.249 &     0.596 &     -19.439 &     10.052 &     TR      
&   -    &   -    \\
\scriptsize{SDSS-J125558.86+302149.2} &    193.9952 &    30.3636 &    0.733 &    0.051 &     1.329 &     0.654 &     -20.065 &     10.292 &     tr*     
&   -   &    -    \\
\scriptsize{SDSS-J130643.54+093911.4} &    196.6814 &    9.6531  &    0.711 &    0.057 &     1.183 &     0.919 &     -19.763 &     10.147 &     -       
&   -   &    -    \\
\scriptsize{SDSS-J131659.28+074326.4} &    199.2470 &    7.7240   &    0.765 &    0.049 &     1.428 &     0.72  &     -19.618 &     10.148 &     -      
&   -   &    -    \\
\scriptsize{SDSS-J133600.37+063133.9} &    204.0015 &    6.5260  &    0.722 &    0.023 &     1.426 &     0.671 &     -19.383 &     10.007 &     TR      
&   -   &    -    \\
\scriptsize{SDSS-J133700.57+432532.1} &    204.2524 &    43.4256 &    0.796 &    0.043 &     0.999 &     0.589 &     -19.472 &     10.124 &     -       
&   -   &    -    \\
\scriptsize{SDSS-J140547.28+151138.3} &    211.4470 &    15.1939 &    0.713 &    0.059 &     1.437 &     0.636 &     -19.467 &     10.031 &     -       
&  -   &    -   \\
\scriptsize{SDSS-J140929.47+000837.2} &    212.3728 &    0.1436  &    0.742 &    0.054 &     1.016 &     0.608 &     -19.614 &     10.121 &     TR      
& -  &  - \\
\scriptsize{SDSS-J144322.25+010553.2} &    220.8427 &    1.0981  &    0.726 &    0.038 &     1.423 &     0.751 &     -20.411 &     10.423 &     TR      
&   -   &   -   \\
\scriptsize{SDSS-J144718.19+581333.3} &    221.8257 &    58.2260  &    0.776 &    0.037 &     1.346 &     0.507 &     -19.883 &     10.266 &     TR     
&   -   &   -   \\
\scriptsize{SDSS-J145403.72+182401.5} &    223.5154 &    18.4004 &    0.814 &    0.057 &     1.475 &     0.596 &     -19.971 &     10.343 &     -       
&   -   &   -   \\
\scriptsize{SDSS-J153235.75+492302.8} &    233.1490 &    49.3842 &    0.785 &    0.052 &     1.409 &     0.54  &     -19.298 &     10.042 &     TR      
&   -   &   -   \\
\scriptsize{SDSS-J161159.99+300251.8} &    242.9999 &    30.0477 &    0.793 &    0.048 &     1.425 &     0.798 &     -19.8   &     10.252 &     -       
&   -   &   -   \\
\scriptsize{SDSS-J163855.82+132327.0} &    249.7326 &    13.3908 &    0.766 &    0.051 &     1.414 &     0.577 &     -19.974 &     10.292 &     -       
&   -   &   -   \\
\scriptsize{SDSS-J170630.27+220003.9} &    256.6261 &    22.0010  &    0.709 &    0.059 &     1.488 &     0.632 &     -19.809 &     10.163 &     TR     
&   -   &   -   \\
\scriptsize{SDSS-J170714.43+652200.2} &    256.8100 &    65.3668 &    0.728 &    0.054 &     1.49  &     0.606 &     -19.819 &     10.188 &     tr*     
&   -   &   -   \\
\scriptsize{SDSS-J221917.33-011113.7} &    334.8222 &    -1.1871 &    0.753 &    0.057 &     1.388 &     0.591 &     -19.619 &     10.135 &     TR      
&   -   &   -   \\
\scriptsize{SDSS-J230751.49+142333.5} &    346.9645 &    14.3926 &    1.001 &    0.037 &     1.276 &     0.822 &     -19.262 &     10.265 &     -       
&   -   &   -   \\
\hline
\end{tabular}
\end{table*}

%% file: coelho-august2013.bbl
\begin{thebibliography}{44}
\expandafter\ifx\csname natexlab\endcsname\relax\def\natexlab#1{#1}\fi

\bibitem[{{Baldwin} {et~al.}(1981){Baldwin}, {Phillips}, \&
  {Terlevich}}]{BPT81}
{Baldwin}, J.~A., {Phillips}, M.~M., \& {Terlevich}, R. 1981, PASP, 93, 5

\bibitem[{{Balogh} {et~al.}(1999){Balogh}, {Morris}, {Yee}, {Carlberg}, \&
  {Ellingson}}]{Balogh99}
{Balogh}, M.~L., {Morris}, S.~L., {Yee}, H.~K.~C., {Carlberg}, R.~G., \&
  {Ellingson}, E. 1999, ApJ, 527, 54

\bibitem[{{Bell}(2008)}]{Bell08}
{Bell}, E.~F. 2008, ApJ, 682, 355

\bibitem[{{Blanton} \& {Roweis}(2007)}]{BlantonRoweis07}
{Blanton}, M.~R. \& {Roweis}, S. 2007, AJ, 133, 734

\bibitem[{{Blanton} {et~al.}(2005){Blanton}, {Schlegel}, {Strauss},
  {Brinkmann}, {Finkbeiner}, {Fukugita}, {Gunn}, {Hogg}, {Ivezi{\'c}}, {Knapp},
  {Lupton}, {Munn}, {Schneider}, {Tegmark}, \& {Zehavi}}]{Blanton05}
{Blanton}, M.~R., {Schlegel}, D.~J., {Strauss}, M.~A., {et~al.} 2005, AJ, 129,
  2562


\bibitem[Bower et al.(2006)]{Bower2006} Bower, R.~G., Benson, 
A.~J., Malbon, R., et al.\ 2006, MNRAS, 370, 645 

\bibitem[{{Budav{\'a}ri} {et~al.}(2009){Budav{\'a}ri}, {Heinis}, {Szalay},
  {Nieto-Santisteban}, {Gupchup}, {Shiao}, {Smith}, {Chang}, {Kauffmann},
  {Morrissey}, {Schiminovich}, {Milliard}, {Wyder}, {Martin}, {Barlow},
  {Seibert}, {Forster}, {Bianchi}, {Donas}, {Friedman}, {Heckman}, {Lee},
  {Madore}, {Neff}, {Rich}, \& {Welsh}}]{Budavari09}
{Budav{\'a}ri}, T., {Heinis}, S., {Szalay}, A.~S., {et~al.} 2009, ApJ, 694,
  1281

\bibitem[{{Cattaneo} {et~al.}(2009){Cattaneo}, {Faber}, {Binney}, {Dekel},
  {Kormendy}, {Mushotzky}, {Babul}, {Best}, {Br{\"u}ggen}, {Fabian}, {Frenk},
  {Khalatyan}, {Netzer}, {Mahdavi}, {Silk}, {Steinmetz}, \&
  {Wisotzki}}]{Cattaneo09}
{Cattaneo}, A., {Faber}, S.~M., {Binney}, J., {et~al.} 2009, Nature, 460, 213

\bibitem[{{Cole} {et~al.}(2000){Cole}, {Lacey}, {Baugh}, \& {Frenk}}]{Cole00}
{Cole}, S., {Lacey}, C.~G., {Baugh}, C.~M., \& {Frenk}, C.~S. 2000, MNRAS,
  319, 168

\bibitem[{{Condon} {et~al.}(1998){Condon}, {Cotton}, {Greisen}, {Yin},
  {Perley}, {Taylor}, \& {Broderick}}]{Condon98}
{Condon}, J.~J., {Cotton}, W.~D., {Greisen}, E.~W., {et~al.} 1998, AJ, 115,
  1693

\bibitem[{{Cortese}(2012)}]{Cortese12}
{Cortese}, L. 2012, A\&A, 543, A132

\bibitem[{{Croton} {et~al.}(2006){Croton}, {Springel}, {White}, {De Lucia},
  {Frenk}, {Gao}, {Jenkins}, {Kauffmann}, {Navarro}, \& {Yoshida}}]{Croton06}
{Croton}, D.~J., {Springel}, V., {White}, S.~D.~M., {et~al.} 2006, MNRAS, 365,
  11

\bibitem[{{Desroches} \& {Ho}(2009)}]{DesrHo09}
{Desroches}, L.-B. \& {Ho}, L.~C. 2009, ApJ, 690, 267

\bibitem[{{Drory} \& {Fisher}(2007)}]{DroryFisher07}
{Drory}, N. \& {Fisher}, D.~B., 2007, ApJ, 664, 640

\bibitem[{{Fisher} \& {Drory}(2010)}]{FisherDrory10}
{Fisher}, D.~B. \& {Drory}, N., 2010, ApJ, 716, 942

\bibitem[{{Falcke} {et~al.}(2000){Falcke}, {Nagar}, {Wilson}, \&
  {Ulvestad}}]{Falcke00}
{Falcke}, H., {Nagar}, N.~M., {Wilson}, A.~S., \& {Ulvestad}, J.~S. 2000, ApJ,
  542, 197

\bibitem[{{Ferrarese} \& {Merritt}(2000)}]{FerraMerr00}
{Ferrarese}, L. \& {Merritt}, D. 2000, ApJl, 539, L9

\bibitem[{{Gebhardt} {et~al.}(2000){Gebhardt}, {Bender}, {Bower}, {Dressler},
  {Faber}, {Filippenko}, {Green}, {Grillmair}, {Ho}, {Kormendy}, {Lauer},
  {Magorrian}, {Pinkney}, {Richstone}, \& {Tremaine}}]{Gebhardt00}
{Gebhardt}, K., {Bender}, R., {Bower}, G., {et~al.} 2000, ApJl, 539, L13

\bibitem[{{Hopkins} {et~al.}(2006){Hopkins}, {Hernquist}, {Cox}, {Di Matteo},
  {Robertson}, \& {Springel}}]{Hopkins06}
{Hopkins}, P.~F., {Hernquist}, L., {Cox}, T.~J., {et~al.} 2006, ApJs, 163, 1

\bibitem[{{Kauffmann} \& {Haehnelt}(2000)}]{KauffHaeh00}
{Kauffmann}, G. \& {Haehnelt}, M. 2000, MNRAS, 311, 576

\bibitem[{{Kauffmann} \& {Heckman}(2009)}]{KauffHeck09}
{Kauffmann}, G. \& {Heckman}, T.~M. 2009, MNRAS, 397, 135

\bibitem[{{Kauffmann} {et~al.}(2003){Kauffmann}, {Heckman}, {Tremonti},
  {Brinchmann}, {Charlot}, {White}, {Ridgway}, {Brinkmann}, {Fukugita}, {Hall},
  {Ivezi{\'c}}, {Richards}, \& {Schneider}}]{Kauff03}
{Kauffmann}, G., {Heckman}, T.~M., {Tremonti}, C., {et~al.} 2003, MNRAS, 346,
  1055

\bibitem[{{Kellermann} {et~al.}(1989){Kellermann}, {Sramek}, {Schmidt},
  {Shaffer}, \& {Green}}]{Keller89}
{Kellermann}, K.~I., {Sramek}, R., {Schmidt}, M., {Shaffer}, D.~B., \& {Green},
  R. 1989, AJ, 98, 1195

\bibitem[{{Kewley} {et~al.}(2001){Kewley}, {Dopita}, {Sutherland}, {Heisler},
  \& {Trevena}}]{Kewley01}
{Kewley}, L.~J., {Dopita}, M.~A., {Sutherland}, R.~S., {Heisler}, C.~A., \&
  {Trevena}, J. 2001, ApJ, 556, 121

\bibitem[{{Kewley} {et~al.}(2006){Kewley}, {Groves}, {Kauffmann}, \&
  {Heckman}}]{Kewley06}
{Kewley}, L.~J., {Groves}, B., {Kauffmann}, G., \& {Heckman}, T. 2006, MNRAS,
  372, 961

\bibitem[{{Kormendy} {et~al.}(2010){Kormendy}, {Drory}, {Bender}, \&
  {Cornell}}]{Kormendy2010}
{Kormendy}, J., {Drory}, N., {Bender}, R., \& {Cornell}, M.~E. 2010, ApJ, 723,
  54

\bibitem[{{Lawrence} {et~al.}(2007)}]{Lawrence07}
{Lawrence}, A. {et~al.}, 2007, MNRAS, 379, 1599

\bibitem[{{Martin} {et~al.}(2007){Martin}, {Wyder}, {Schiminovich}, {Barlow},
  {Forster}, {Friedman}, {Morrissey}, {Neff}, {Seibert}, {Small}, {Welsh},
  {Bianchi}, {Donas}, {Heckman}, {Lee}, {Madore}, {Milliard}, {Rich}, {Szalay},
  \& {Yi}}]{Martin07}
{Martin}, D.~C., {Wyder}, T.~K., {Schiminovich}, D., {et~al.} 2007, ApJs, 173,
  342

\bibitem[Masters et al.(2010)]{Masters2010} Masters, K.~L., Mosleh, 
M., Romer, A.~K., et al.\ 2010, MNRAS, 405, 783 

\bibitem[{{McAlpine} {et~al.}(2011){McAlpine}, {Satyapal}, {Gliozzi}, {Cheung},
  {Sambruna}, \& {Eracleous}}]{McAlpine11}
{McAlpine}, W., {Satyapal}, S., {Gliozzi}, M., {et~al.} 2011, ApJ, 728, 25

\bibitem[{{Nagar} {et~al.}(2005){Nagar}, {Falcke}, \& {Wilson}}]{Nagar05}
{Nagar}, N.~M., {Falcke}, H., \& {Wilson}, A.~S. 2005, A\&A, 435, 521

\bibitem[{{O'Donnell}(1994)}]{ODonnell94}
{O'Donnell}, J.~E. 1994, ApJ, 422, 158

\bibitem[{{Peng} {et~al.}(2010){Peng}, {Ho}, {Impey}, \& {Rix}}]{Peng2010}
{Peng}, C.~Y., {Ho}, L.~C., {Impey}, C.~D., \& {Rix}, H.-W. 2010, AJ, 139,
  2097

\bibitem[{{Peterson} {et~al.}(2005){Peterson}, {Bentz}, {Desroches},
  {Filippenko}, {Ho}, {Kaspi}, {Laor}, {Maoz}, {Moran}, {Pogge}, \&
  {Quillen}}]{Peterson2005}
{Peterson}, B.~M., {Bentz}, M.~C., {Desroches}, L.-B., {et~al.} 2005, ApJ,
  632, 799

\bibitem[{{Phillips} {et~al.}(1996)}]{Phillips96}
{Phillips}, A.~C., {Illingworth}, G.~D., {MacKenty}, J.~W., {Franx}, M., 1996, AJ, 111, 1566

\bibitem[{{Rieke} {et~al.}(2009){Rieke}, {Alonso-Herrero}, {Weiner},
  {P{\'e}rez-Gonz{\'a}lez}, {Blaylock}, {Donley}, \& {Marcillac}}]{Rieke09}
{Rieke}, G.~H., {Alonso-Herrero}, A., {Weiner}, B.~J., {et~al.} 2009, ApJ,
  692, 556

\bibitem[{{Sachdeva} (2013)}]{Sachdeva13}
{Sachdeva}, S., 2013, arXiv:1307.6184 (MNRAS accepted)

\bibitem[{{Satyapal} {et~al.}(2009){Satyapal}, {B{\"o}ker}, {Mcalpine},
  {Gliozzi}, {Abel}, \& {Heckman}}]{Satyapal09}
{Satyapal}, S., {B{\"o}ker}, T., {Mcalpine}, W., {et~al.} 2009, ApJ, 704, 439

\bibitem[{{Satyapal} {et~al.}(2008){Satyapal}, {Vega}, {Dudik}, {Abel}, \&
  {Heckman}}]{Satyapal08}
{Satyapal}, S., {Vega}, D., {Dudik}, R.~P., {Abel}, N.~P., \& {Heckman}, T.
  2008, ApJ, 677, 926

\bibitem[{{Satyapal} {et~al.}(2007){Satyapal}, {Vega}, {Heckman}, {O'Halloran},
  \& {Dudik}}]{Satyapal07}
{Satyapal}, S., {Vega}, D., {Heckman}, T., {O'Halloran}, B., \& {Dudik}, R.
  2007, ApJl, 663, L9

\bibitem[{{Schawinski} {et~al.}(2007){Schawinski}, {Thomas}, {Sarzi},
  {Maraston}, {Kaviraj}, {Joo}, {Yi}, \& {Silk}}]{Schawinski07}
{Schawinski}, K., {Thomas}, D., {Sarzi}, M., {et~al.} 2007, MNRAS, 382, 1415

\bibitem[{{Schlegel} {et~al.}(1998){Schlegel}, {Finkbeiner}, \&
  {Davis}}]{Schlegel98}
{Schlegel}, D.~J., {Finkbeiner}, D.~P., \& {Davis}, M. 1998, ApJ, 500, 525

\bibitem[{{Secrest} {et~al.}(2012){Secrest}, {Satyapal}, {Gliozzi}, {Cheung},
  {Seth}, \& {B{\"o}ker}}]{Secrest12}
{Secrest}, N.~J., {Satyapal}, S., {Gliozzi}, M., {et~al.} 2012, ApJ, 753, 38

\bibitem[{{Shields} {et~al.}(2008){Shields}, {Walcher}, {B{\"o}ker}, {Ho},
  {Rix}, \& {van der Marel}}]{Schields2008}
{Shields}, J.~C., {Walcher}, C.~J., {B{\"o}ker}, T., {et~al.} 2008, ApJ, 682,
  104

\bibitem[Simmons et al.(2013)]{Simmons2013} Simmons, B.~D., 
Lintott, C., Schawinski, K., et al.\ 2013, MNRAS, 444 


\bibitem[{{Stasi{\'n}ska} {et~al.}(2006){Stasi{\'n}ska}, {Cid Fernandes},
  {Mateus}, {Sodr{\'e}}, \& {Asari}}]{Stasi06}
{Stasi{\'n}ska}, G., {Cid Fernandes}, R., {Mateus}, A., {Sodr{\'e}}, L., \&
  {Asari}, N.~V. 2006, MNRAS, 371, 972

\bibitem[{{Tremonti} {et~al.}(2004){Tremonti}, {Heckman}, {Kauffmann},
  {Brinchmann}, {Charlot}, {White}, {Seibert}, {Peng}, {Schlegel}, {Uomoto},
  {Fukugita}, \& {Brinkmann}}]{Tremonti04}
{Tremonti}, C.~A., {Heckman}, T.~M., {Kauffmann}, G., {et~al.} 2004, ApJ, 613,
  898

\bibitem[{{Tremonti} {et~al.}(2007){Tremonti}, {Moustakas}, \&
  {Diamond-Stanic}}]{Tremonti07}
{Tremonti}, C.~A., {Moustakas}, J., \& {Diamond-Stanic}, A.~M. 2007, ApJl,
  663, L77

\bibitem[{{Vincent} \& {Ryden}(2005)}]{Vincent05}
{Vincent}, R.~A. \& {Ryden}, B.~S. 2005, ApJ, 623, 137

\bibitem[{{White} {et~al.}(1997){White}, {Becker}, {Helfand}, \& {Gregg}}]{W97}
{White}, R.~L., {Becker}, R.~H., {Helfand}, D.~J., \& {Gregg}, M.~D. 1997,
  ApJ, 475, 479

\bibitem[{{Williams} {et~al.}(2009){Williams}, {Quadri}, {Franx}, {van Dokkum},
  \& {Labb{\'e}}}]{Williams09}
{Williams}, R.~J., {Quadri}, R.~F., {Franx}, M., {van Dokkum}, P., \&
  {Labb{\'e}}, I. 2009, ApJ, 691, 1879

\bibitem[{{Yan} \& {Blanton}(2012)}]{Yan2012}
{Yan}, R. \& {Blanton}, M.~R. 2012, ApJ, 747, 61

\end{thebibliography}
